\documentclass[a4paper,11pt]{article}

\usepackage[margin=1in]{geometry}

\usepackage{graphicx,amsmath,amssymb}
\usepackage{stackengine}

\usepackage[utf8]{inputenc}
\usepackage[english]{babel}
\usepackage{csquotes}

\usepackage{aas_macros}

\usepackage{xcolor}
\usepackage{hyperref}

\hypersetup{
  colorlinks,
  citecolor=blue,
  linkcolor=red,
  urlcolor=olive}

\newcommand{\planck}{\emph{Planck}}
\newcommand{\commander}{\texttt{Commander}}

\newcommand{\npipe}{\texttt{NPIPE}}
\newcommand{\healpix}{\texttt{HEALPix}}
\newcommand{\isap}{\texttt{iSAP}}

\newcommand*{\Comb}[2]{{}^{#1}C_{#2}}%

\title{Examining statistical isotropy of CMB low multipoles from \planck\ PR4 data}

\author{Pavan K. Aluri\footnote{pavanaluri.phy@iitbhu.ac.in}, Sanjeet Kumar Patel\footnote{sanjeetkumarpatel.rs.phy18@iitbhu.ac.in }}
\date{}

\begin{document}

\maketitle

\centerline{Department of Physics, Indian Institute of Technology (BHU), Varanasi - 221005, India}

\begin{abstract}
Low multipoles ($l$)  in cosmic microwave background (CMB) temperature anisotropies have shown some `peculiarities' when examined since the release of the full sky CMB maps, using a variety of tests. In this paper, we concern ourselves with the very first peculiarities seen in CMB data viz., a breakdown of statistical isotropy in the form of axiality and planarity of these low-$l$ modes, and preferred alignments among them. We scrutinize the latest CMB data from ESA's \planck\ mission, PR4, to evaluate the current status of these deviations. We employ the Power tensor method which allows an invariant characterization of the distribution of power in a given multipole, and apply it to probe the first sixty
multipoles i.e., $l=2$ to 61. We find that there are significant number of modes that are intrinsically anisotropic with a cumulative probability of $0.3\%$. However since the planarity study reveals that those modes that are unusually planar are subset of these anisotropic modes, we conclude that they may not be intrinsically planar. The quadrupole is still well aligned with the octopole. Besides, $l=3$, higher
multipoles aligned with quadrupole are found to be insignificant. Interestingly, the collective alignment axis of the first sixty multipoles is found to be broadly closer to the axis of dipole, quadrupole, octopole and other modes aligned with $l=2$.
\end{abstract}


\section{Introduction}
\emph{Cosmological principle} which is the foundation of modern cosmology asserts that the universe is homogeneous and isotropic on very large scales. With the advent of precision cosmology as a result of full sky mapping of cosmic microwave background (CMB) anisotropies, particularly starting from NASA’s Wilkinson Microwave Anisotropy Probe’s (WMAP) first year data release, these implications, specifically the isotropy of CMB, was subjected to a plethora of tests. This exercise by the cosmology community resulted in identifying some peculiarities of CMB low multipoles that have come to be called \emph{anomalies} (for an overview of these anomalies see, for example, \cite{Schwarz16,Bull16,Muir18}). These were found to be present in the CMB data from the latest ESA's \planck\ full sky mission also, that succeeded the WMAP mission \cite{WMAP7yrAnom,WMAP9yrFinalmaps,Planck13isostat,Planck15isostat,Planck18isostat}.

One of the first few anomalies that received significant attention is the existence of preferred alignments among low multipoles \cite{Tegmark04,Schwarz04,Copi04,RalstonJain04,Slosar04,Bielewicz04,Bielewicz05,Land05,Tegmark06}.
These large angular scale modes were found to be anomalously aligned or highly planar. In this paper, we assess the status of these alignments using the latest CMB temperature anisotropies
map from \planck\ public release 4 (PR4) obtained using the \commander\ cleaning procedure \cite{Eriksen04,Eriksen08}.

\planck\ PR4 maps were produced using the \npipe\ map making pipeline \cite{Planck2020}. Some of the statistical tools to examine such preferred alignments proposed in the literature thus far are, for example, the angular momentum maximization method \cite{Tegmark04,Tegmark06}, Maxwell's multipole vectors \cite{Schwarz04,Copi04}, and the Power tensor method \cite{RalstonJain04,Samal08}. In this paper, we will be using the Power tensor method. More details on the relationship between our chosen method and the other methods can be found in Ref. \cite{RalstonJain04,Samal08}.

Rest of the paper is organized as follows. First, the Power tensor method and associated statistics are briefly reviewed. Then the data and simulations used in the current study to assess the significance of our statistics are described. Thereafter, our results and analysis are presented, followed by discussion and conclusions.

\section{Statistics used}
In this section, we describe the statistics used for the analysis of CMB maps viz., the Power tensor (PT) and the Alignment tensor (AT) developed in Ref.~\cite{RalstonJain04,Samal08}. However in what follows we will be presenting their slightly altered definitions as used in Ref.~\cite{Aluri17}. This methodology was also applied to $E$-mode polarization of CMB from \emph{Planck} 2015 full mission data in Ref.~\cite{Rath18}, however with inconclusive results owing to its low signal to noise ratio and foreground residual biases.

\subsection{Power tensor}

CMB anisotropies are conventionally expanded in terms of spherical harmonics, $Y_{lm}(\hat{n})$, as
\begin{equation}
\Delta T(\hat{n}) = T(\hat{n}) - T_0\, = \sum_{l=1}^\infty \sum_{m=-l}^{+l} a_{lm} Y_{lm}(\hat{n})\,,
\end{equation}
where $\Delta T(\hat{n})$ are the temperature fluctuations, around the mean CMB sky temperature of $T_0\approx2.7255$~K~\cite{Fixsen09}, in some direction $\hat{n}=(\theta,\phi)$ on the celestial sphere. The Power tensor is a quadratic estimator in terms of spherical harmonic coefficients, $a_{lm}$, and is defined as,
\begin{equation}
A_{ij}(l) = \frac{1}{l(l+1)(2l+1)} \sum_{m m' m''} a_{lm}^* J^i_{mm'} J^j_{m'm''} a_{lm''}\,,
\label{eq:pt}
\end{equation}
where $J^i_{mm'}$, for $i=x,y,z$, are the $(2l+1)\times (2l+1)$ angular momentum matrices for a
particular `$l$'. Thus, the Power tensor is a $3\times 3$ matrix, and the prefactor ensures that the
ensemble average of the Power tensor $\langle A_{ij} \rangle = \delta_{ij}\,C_l/3$ and it's trace
given by $\sum_i \langle A_{ii} \rangle = C_l$. This follows from the fact that the two point
angular correlation function of CMB anisotropies is given by
$\langle a_{lm} a^*_{l'm'} \rangle = C_l \delta_{ll'}\delta_{mm'}$, when statistical isotropy holds.

The Power tensor essentially maps a multipole onto an ellipsoid, with the three orthogonal axes given by its three eigenvectors defining an invariant frame. It's three eigenvalues indicate the amount of power directed along each axis. Let $\Lambda_\alpha$, $\alpha=1,2,3$ be the eigenvalues, and 
$e^\alpha_k$ be the corresponding eigenvectors, ${\bf e}^\alpha$, with components indexed by `$k$'.
Further, let the normalized eigenvalues be denoted by $\lambda_\alpha = \Lambda_\alpha/(\sum_\beta \Lambda_\beta)$. Using these (normalized) eigenvalues, one can define what is called a \emph{Power entropy} as,
\begin{equation}
S(l) = -\sum_{\alpha=1}^3 \lambda_\alpha \log(\lambda_\alpha)\,,
\label{eq:pe}
\end{equation}
where `$\log$' in the above equation is the natural logarithm of the normalized eigenvalues.

Depending on how much power ($\lambda_\alpha$) is distributed along each eigenvector,
one can characterize the nature of a multipole.
If all the power is contained in one of the eigenvalues of $A_{ij}$, say $\Lambda_3$, then
$\lambda_3 \rightarrow 1$, while $\lambda_1$, $\lambda_2 \rightarrow 0$, and the
Power entropy, $S \rightarrow 0$. Thus the ellipsoid represented by the Power tensor for
a particular multipole is maximally deformed along an axis. In the case of statistical isotropy, the
three eigenvalues will be equal i.e., $\lambda_\alpha=1/3$, for $\alpha=1,2,3$, and the Power entropy
$S=\log(3)\approx 1.0986$ represents maximal isotropy of a multipole `$l$'.
Note that, the eigenvectors of the Power tensor are headless, meaning they only represent axes and not directed vectors. We can now associate a \emph{preferred axis} with a multipole `$l$', as the one along which more power is directed (i.e., eigenvector corresponding to the largest eigenvalue). We refer to this axis as principal eigenvector (PEV).

The eigenvalues and eigenvectors, and thus PEVs, provide independent information about the nature of the multipole. The eigenvalues can be invariantly combined using Power entropy to test the level of isotropy violation in a particular multipole. Independent of the Power entropy, the principal eigenvectors can be used to compare alignments among various multipoles. Let us denote the PEV of a multipole as $\tilde{\bf e}_l$.
In the case of statistical isotropy, the orientation of PEVs are equi-spaced with respect to each other.

To characterize the same i.e., the level of alignment between any two multipoles, let us define the statistic $1 - \cos\theta_{ll'} = 1- \tilde{\bf e}_l \cdot \tilde{\bf e}_{l'}$, where $\theta_{ll'}$ is the angle between the two PEVs $\tilde{\bf e}_l$ and $\tilde{\bf e}_{l'}$ corresponding to $l$ and $l'$. For isotropically distributed axes, the statistic ``$x=1 - \cos\theta_{ll'}$'' has a probability density, $P(x)$, that is uniform in the interval $x=[0,1]$. Since the PEVs are headless, the range of alignments possible between them are $\theta_{ll'} = [0^\circ,90^\circ]$ (degrees). Thus, ``$\cos\theta_{ll'}$'' and correspondingly ``$1-\cos\theta_{ll'}$'' fall in the interval [0,1].

\subsection{Alignment tensor}
In order to probe the nature of alignments among a set of multipoles or over a chosen range of multipoles i.e., to extract a common alignment axis for those modes of our interest, we use
what is called an \emph{Alignment tensor} that is defined as
\begin{equation}
X_{ij} = \frac{1}{N_l}\sum_l \tilde{e}^i_l \tilde{e}^j_l\,,
\label{eq:at}
\end{equation}
where $\tilde{\bf e}_l$ is the principal eigenvector (PEV) of a multipole `$l$'.
There is a trivial multiplicative factor that we get as a result of the summation in calculating this Alignment tensor which is the number of multipoles used, $N_l$, that can be removed.
Thus, let $\zeta_\alpha$ ($\alpha=1,2,3$) be the normalized eigenvalues, and ${\bf f}_\alpha$
be the normalized eigenvectors of the Alignment tensor. Further,
let $\tilde{\bf f}$ be the eigenvector corresponding to the largest eigenvalue, say, $\zeta_3$,
after sorting $\zeta_\alpha$ in ascending order. 
Similar to the Power tensor, these normalized eigenvalues, $\zeta_\alpha$, can be used to compute what
is called \emph{Alignment entropy} that is defined as,
\begin{equation}
S_X = -\sum_{\alpha=1}^3 \zeta_\alpha \log(\zeta_\alpha)\,,
\label{eq:ae}
\end{equation}
that also varies in the range $[0,\log(3)]$,
where the lower and higher ends of this interval represent perfectly aligned axes (PEVs)
and perfectly isotropically distributed PEVs, respectively.

The eigenvector corresponding to the largest eigenvalue of the Alignment tensor can be taken to represent a `preferred' axis for the set of multipole PEVs used to compute the Alignment tensor. However, it will be meaningful only when the largest eigenvalue of the Alignment tensor has high significance as inferred from complementary simulations. On the other hand, if the smallest eigenvalue is anomalously small, then it would indicate that the PEVs are predominantly lying in a plane defined by the corresponding eigenvector, which is normal to that plane

For more details on Power tensor, Alignment tensor and related statistics described here, the reader may refer to Ref.~\cite{RalstonJain04,Samal08,Aluri11}.

\section{Data and complementary simulations}\label{sec:data-sim}
In this work, we use the latest full sky CMB temperature anisotropy maps from European Space Agency's (ESA)
\planck\  mission\footnote{\url{https://sci.esa.int/web/planck}}, processed using
the \npipe\ pipeline \cite{Planck2020} and cleaned using the \commander\
algorithm \cite{Eriksen04,Eriksen08}. These are made available through the public data release 4
(PR4) of \planck\ mission at Planck legacy archive\footnote{\url{https://www.cosmos.esa.int/web/planck}} (PLA). This cleaned CMB map is provided at a \healpix\footnote{\url{https://healpix.sourceforge.io/}} pixel resolution of $N_{side}=4096$ smoothed with a Gaussian beam
of $FWHM=5'$ (arcmin).

However, since we are interested in analyzing low multipoles, this map is
downgraded to a resolution of \healpix\ $N_{side}=256$ smoothed to have
an effective Gaussian beam of $FWHM=40'$. The \npipe\ \commander\ CMB map
is produced with the dipole mode ($l=1$) retained, unlike
the CMB maps from previous public data releases of \planck\ \cite{Planck2020}.
We proceed as follows to downgrade the cleaned CMB map. First, the $l=1$ multipole is fitted out from
the \commander\ CMB map outside the galactic mask with an additional symmetric cut of
$\cos(b)=\pm 0.1$ about the galactic plane (where `$b$' is the galactic co-latitude)
using the \verb+remove_dipole+ functionality of \healpix.
Once the (monopole and) dipole are fitted out, the spherical harmonic co-efficients are
extracted from this higher $N_{side}$ map, convolved with appropriate Gaussian beam ($b_l$) and
pixel window ($p_l$) functions as
\begin{equation}
a_{lm}^{out} = \frac{b_l^{40'}p_l^{256}}{b_l^{5'}p_l^{4096}} a_{lm}^{in}\,,
\label{eq:con-decon}
\end{equation}
and converted to an $N_{side}=256$ map with a Gaussian beam resolution of $FWHM=40'$.
Now the residual foregrounds are inpainted using the \isap\ software\footnote{\url{http://www.cosmostat.org/software/isap}} package that fills potentially contaminated regions as
defined by a foreground mask through a Sparse inpainting technique while still preserving the
statistical properties with the rest of the unmasked sky. More details on the CMB maps and masks
used in this analysis are given in Appendix~\ref{apdx1}.


Complementary simulations to the \planck\ PR4 \npipe\ \commander\ CMB maps are also provided with
this public release. These mock data, referred to as Full Focal Plane (FFP) simulations incorporating
all the detector characteristics and processed using the same pipeline as the data, are made available
on \texttt{NERSC}\footnote{\url{https://crd.lbl.gov/cmb-data}} computing facility. Currently, 100 simulations complementary to the \npipe\ \commander\
temperature anisotropy map are available. These are also provided at the same \healpix\ $N_{side}=4096$, as the data, with a Gaussian beam smoothing of $FWHM=5'$.

In addition to these 100 simulations, we also generated 5000 ideal CMB realizations based
on the theoretical temperature power spectrum ($C_l$) derived from
\texttt{CAMB}\footnote{\url{https://camb.info/}} using the latest
\planck\ estimated cosmological parameters \cite{planck2018cosmopar}. These are generated at
\healpix\ $N_{side}=256$ with a Gaussian beam resolution of $FWHM=40'$. Since, noise is negligible at low multipoles, no further (random) noise component is added to these maps. The Power tensor statistic is also applied on these 5000 CMB realizations to get a better quantification of the
statistics than we get when using only 100 \npipe\ \commander\ simulations provided by \planck\ collaboration. The suitability of ideal CMB realizations for our present purpose is studied,
for example, in Ref.~\cite{Muir18,Aluri11}.

In the analysis that is to follow, we show the significance of various statistics described earlier, as estimated from both the 100 \planck\ 2020 FFP simulations and the 5000 ideal CMB realizations that we generated. We find that the estimates compare fairly well, despite the difference in the number of simulations used and the ideal nature of the second set of simulations we generated.

\section{Analysis and results}
\subsection{Probing axial and planar modes with Power tensor}
Power entropy, $S(l)$, allows for combining the eigenvalues of
the Power tensor so as to characterize the breakdown of isotropy
in a particular multipole `$l$'. If $S\rightarrow 0$, for a multipole, then that
mode is highly axial with a preferred axis given by the principal eigenvector (PEV) of the Power tensor. On the otherhand, if $S \rightarrow \log(3)$, then statistical isotropy is preserved.

\begin{figure}
\centering
\includegraphics[width=0.81\textwidth]{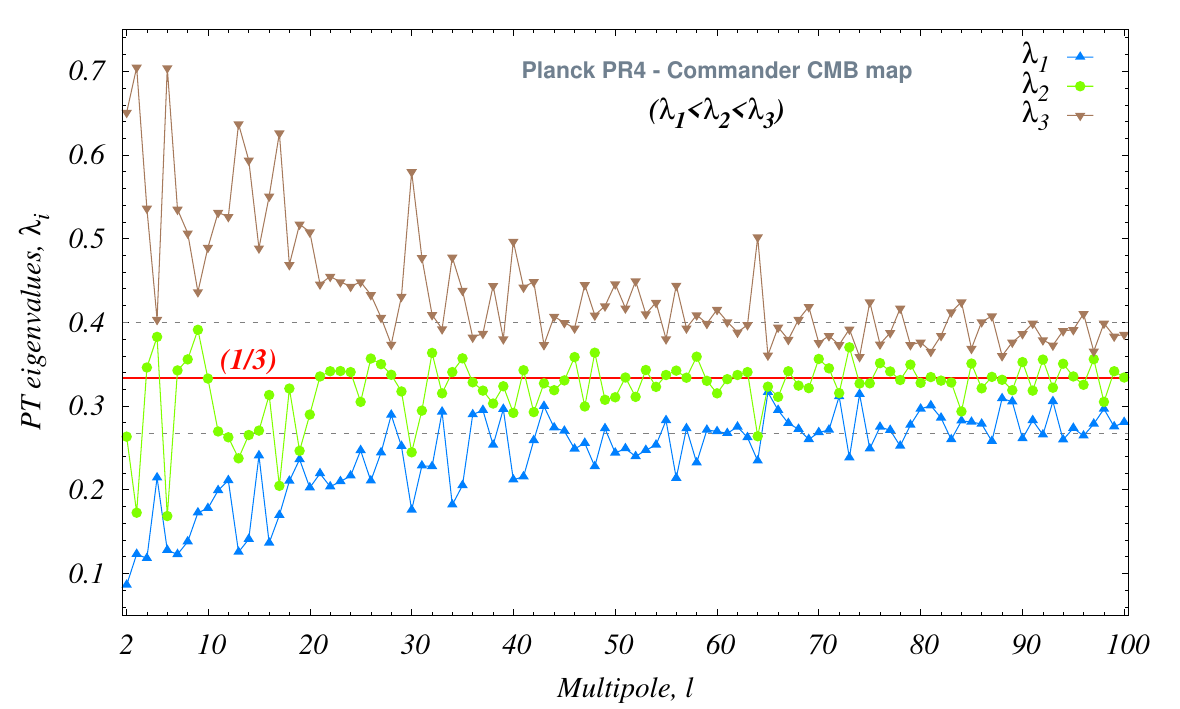}
\caption{Eigenvalues of Power tensor for each `$l$' in the CMB map from \planck\ PR4
         data obtained using \commander\ method. The horizontal red line is the
         expectation value of the eigenvalues when isotropy holds viz.,
         $\langle \lambda_i \rangle = 1/3$, and the two dashed lines denote the $\pm20\%$
         deviation around this expectation i.e., $\pm 1/5^{th}\times 1/3=\pm1/15$.}
\label{fig:pt_eval}
\end{figure}

\begin{figure}
\centering
\includegraphics[width=0.81\textwidth]{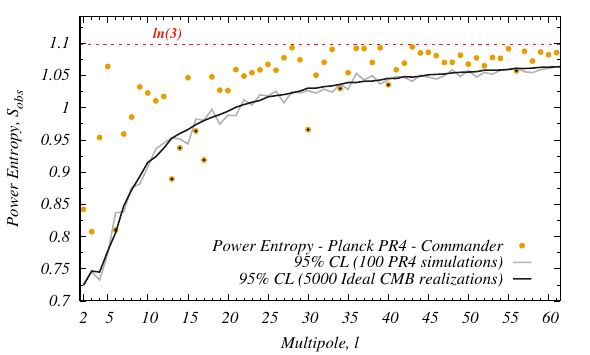}\\
\includegraphics[width=0.81\textwidth]{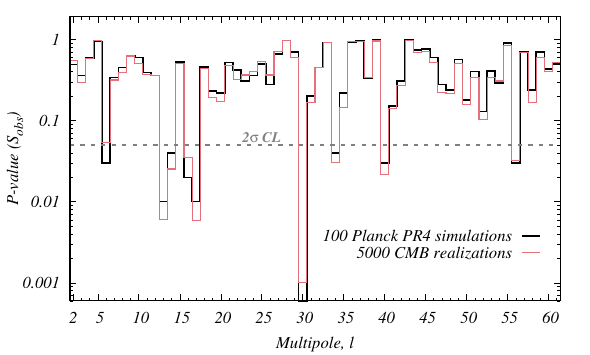}
\caption{\emph{Top :} Power entropy with 95\% confidence level derived
         from 100 \planck\ PR4 \commander\ simulations and 5000 ideal CMB
         realizations based on the theoretical power spectrum from \planck\
         2018 data release. \emph{Bottom :} $p$-value plot of the Power entropy
         statistic obtained by comparing the data statistic from \planck\ PR4
         \commander\ CMB map with the same from simulations. Here also the
         two curves correspond to $p$-values obtained using two sets of simulation
         ensembles.}
\label{fig:pe}
\end{figure}

The three (normalized) eigenvalues of the Power tensor at each multipole for the downgraded \npipe\ \commander\ 2020 CMB map are shown in Fig.~[\ref{fig:pt_eval}], up to $l=100$.
However, we limit rest of the analysis presented in this study up to the first sixty multipoles only
i.e., $l=$2 to 61.
The eigenvalues are sorted in ascending order and labeled $\lambda_1$, $\lambda_2$ and
$\lambda_3$ accordingly.
The expectation value of these normalized eigenvalues i.e., $\langle \lambda_i \rangle = 1/3$,
when isotropy holds, is also plotted in the same figure as a horizontal red line, further illustrating a $\pm 20\%$ (i.e., $\pm 1/15$) deviation from this expected value of  $\langle \lambda_i \rangle = 1/3$
as two dashed lines around it. One can readily see that the low multipoles show an apparent deviation
from $1/3$.
But this visual discrepancy will be quantified later using the Power entropy statistic using simulations.

In \emph{top} panel of Fig.~[\ref{fig:pe}], the Power entropy, $S(l)$, thus computed using these normalized eigenvalues for different multipoles, is shown as yellow points for the \npipe\ \commander\ cleaned CMB map synthesized at \healpix\ $N_{side}=256$
(that is obtained from the original map provided by \planck\ team at $N_{side}=4096$
as described in the preceding section).
The grey and black color curves in the \emph{top} panel of Fig.~[\ref{fig:pe}] are the 95\%
confidence limits (CL) for the
Power entropy at each $l$, as obtained from 100 \planck\ provided simulations and 5000 ideal
CMB realizations based on best fit theoretical power spectrum corresponding to the latest
cosmological parameter estimates from \planck. In spite of the different number of mock maps comprising the two sets of simulations, the two curves agree well with each other. Those multipoles whose Power entropy is outside the 95\% confidence level (as defined by any of the two sets of simulations) are highlighted with a black dot inside the yellow dots. They are further listed in Table~\ref{tab:anom-pe}.

In the \emph{bottom} panel of Fig.~[\ref{fig:pe}], the $p$-values for Power entropy
of all the first sixty
multipoles i.e., $l=[2,61]$ are shown as a function of `$l$'. The $2\sigma$ (95\%) confidence level is also indicated. This is an alternate illustration of the top panel in the same figure, with $p$-values more readily discernable corresponding to each multipole.

\begin{table}[!ht]
\centering
\begin{tabular}{c c}
\hline
Multipole, $l$ & $p$-value of $S_{\rm obs}(l)$ \\

\hline

6  & ~0.03 (0.0534) \\
13 & ~0.01 (0.0060) \\
14 & ~0.04 (0.0254)\\
16 & ~0.02 (0.0354) \\
17 & ~0.01 (0.0058) \\
30 & $<0.01$ (0.0010) \\
34 & ~0.04 (0.0300) \\
40 & ~0.03 (0.0218) \\
56 & ~0.03 (0.0320)\\

\hline
\end{tabular}
\caption{List of multipoles whose Power entropy $p$-value is anomalous
with a random chance occurrence probability of $\leq 0.05$. The $p$-values mentioned
principally are estimated from 100 \planck\ provided FFP 2020 simulations, while those
in the braces are obtained using 5000 ideal CMB realizations. The multipole, $l$,
for which the $p$-value is indicated as `$<0.01$' corresponds to the case where none
of the 100 \planck\ FFP simulations have a Power entropy, $S(l)$, smaller than 
$S_{\rm obs}(l)$ from data.}
\label{tab:anom-pe}
\end{table}

%
%

\begin{figure}
\centering
\includegraphics[width=0.81\textwidth]{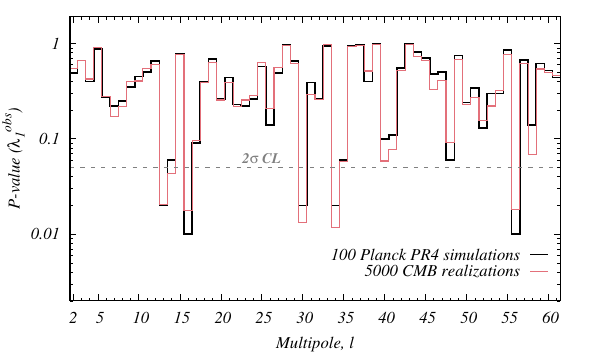}
\caption{$p$-value plot of the smallest eigenvalue of Power tensor, $\lambda_1^{\rm obs}$,
         for the first sixty multipoles of the CMB map derived using the \planck\ \npipe\
         processing pipeline (PR4) and cleaned using the \commander\ method. The
         $2\sigma$ (95\%) level is also indicated by a gray horizontal line. The curves
         correspond to the two different simulation sets used viz., 100 \planck\ like
         simulations provided by \planck\ collaboration and 5000 mock CMB maps based
         on best fit theoretical power spectrum from \planck\ derived latest cosmological
         parameters.}
\label{fig:pt-e1-pval}
\end{figure}

Now, in order to probe the planarity of a CMB mode, we compare the smallest eigenvalue of the Power tensor of a multipole from data with the one that is obtained from the two sets of simulations being used. There are also other statistics, for example, based on the angular momentum maximization of a CMB mode \cite{Tegmark04}, that can be used to probe the planarity of CMB multipoles. However, here we proceed with comparing the smallest eigenvalue of the Power tensor with the same from simulations. The $p$-value plot of the smallest eigenvalue of the Power tensor corresponding to a multipole from data in comparison to simulations is shown in Fig.~[\ref{fig:pt-e1-pval}] along with the $2\sigma$ level. Multipoles
that are found to have a $p$-value $\leq 0.05$ are $l=$13, 14, 16, 30, 34, and 56.
These multipoles and their corresponding $p$-values are listed in Table~\ref{tab:anom-e1}.
However we note that, all these modes also have an anomalous Power entropy (see Table~\ref{tab:anom-pe}).
Thus the anomalous nature of the smallest eigenvalue of the Power tensor of these modes is essentially due to the anomalously large largest eigenvalue of the Power tensor of the same multipoles. Thus we deduce that there are no truly planar modes in the CMB map analyzed in this study.

\begin{table}[!ht]
\centering
\begin{tabular}{c c}
\hline
Multipole, $l$ & $p$-value of $\lambda_1^{\rm obs}$ \\

\hline

13 & ~0.02 (0.1252) \\
14 & ~0.06 (0.1475)\\
16 & ~0.01 (0.1788) \\
30 & ~0.02 (0.3245) \\
34 & ~0.02 (0.3254) \\
56 & ~0.01 (0.3289)\\

\hline
\end{tabular}
\caption{Same as Table~\ref{tab:anom-pe} but for multipoles whose $p$-value of the smallest
         eigenvalue of the Power tensor is anomalous with a random chance occurrence probability of
         $\leq 0.05$.}
\label{tab:anom-e1}
\end{table}

\subsection{Alignments of higher order multipoles with the quadrupole}
\label{sec:pt-algn-2l}
In the cleaned CMB map obtained from first year data of WMAP's observations, it was found that the $l=2,3$ modes are very well aligned along 
the direction of the CMB kinetic dipole \cite{Tegmark04,RalstonJain04}.
It also persisted to be seen in the CMB maps from an entirely different full sky CMB mission i.e., ESA's \planck\ probe. This was interpreted to be the case for a preferred direction for our universe on large scales. Similar preferred alignments were also found using the Maxwell's multipole vectors (MMVs) corresponding to the low-$l$'s \cite{Copi04,Bielewicz04}.
Recall that for each multipole, $l$, there will be `$l$' unit norm headless vectors from an MMV decomposition of spherical harmonic co-efficients. Thus, various combinations of these `$l$' MMVs for each multipole, and also
by combining MMVs from different `$l$', some preferred alignments were found in the data. But, for higher multipoles, these combinations become excessively large in number.

So, first, we test the status of anomalous alignment between the quadrupole ($l=2$) and
the octopole ($l=3$) modes using the PEVs. As mentioned before, while the Power entropy calculated from the eigenvalues of the Power tensor sheds light on the axiality of a particular multipole, the eigenvectors contain independent information along which the power (magnitude of eigenvalues of the Power tensor) is distributed which define an invariant orthonormal frame. Thus the PEVs, viz., the eigenvector corresponding to the largest eigenvalue of the Power tensor of a multipole, can be used to test for any preferred alignments among various multipoles. Here we first look at the alignment between the $l=2,3$ multipoles, and later analyze alignments of higher multipoles with the quadrupole.

The $p$-values obtained using the two simulation ensembles viz., \planck\ like and ideal
CMB realizations, are $0.08$ and $0.035$, respectively, for the observed level of alignment
in the \npipe\ \commander\ derived CMB temperature anisotropies map between $l=2,3$ modes.
They are aligned at a mere angular separation of
$\theta_{23}=\cos^{-1}(\tilde{\bf e}_2\cdot \tilde{\bf e}_3) \approx 14.8^\circ$ (degrees).
We recall that, using the Power tensor method on WMAP's first year ILC map, this alignment was
found to be $\approx 9.94^\circ$ with a random chance occurance probability
of $p=0.02$~\cite{RalstonJain04}. A more elaborate analysis using PT with realistic simulations
revealed that these modes are aligned at a mere angular separation of
$\approx 5.97^\circ$~\cite{Samal08}. In WMAP fifth and seventh year ILC maps, the angular separation
further reduced to $\approx 1.95^\circ$ and $0.6^\circ$ respectively, with a corresponding
increase in it's statistical significance. It is important to note that a foreground bias correction map was subtracted from ILC maps after cleaning, using appropriate simulations~\cite{Hinshaw2007}. This was also found to be the case with cleaned CMB maps
obtained using harmonic space ILC technique as discussed in Ref.~\cite{Aluri11}. Thus, foreground
residuals have a tendency to misalign the multipole PEVs. So this observed alignment
between $l=2,3$ in Planck's PR4 CMB map (and perhaps also between any two multipole PEVs)
in which no such foreground residual bias correction was applied, may be treated as a conservative estimate of the level of alignment between them, and is likely to be more pronounced in light of these previous studies.

\begin{figure}[hb]
\centering
\includegraphics[width=0.81\textwidth]{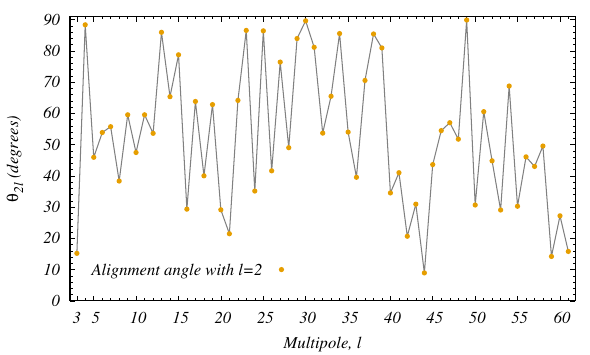}\\
\includegraphics[width=0.81\textwidth]{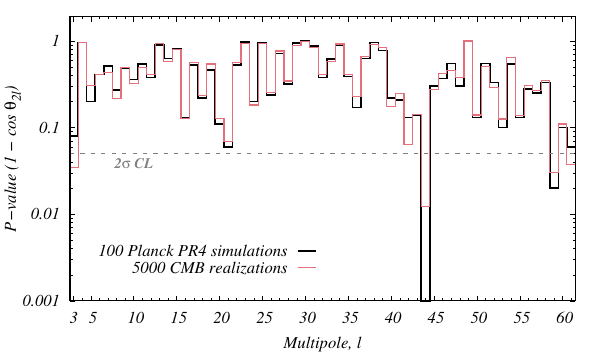}
\caption{\emph{Top :} Alignment angle between CMB quadrupole ($l=2$) and higher multipoles
         upto $l_{max}=61$ in \planck\ PR4 \commander\ map.
         \emph{Bottom :} $p$-values of the observed alignment statistic ``$1-\cos\theta_{2l}$''
         characterizing the alignment between quadrupole and rest of the multipoles
         $l=3$ to 61.}
\label{fig:algn-l2ell}
\end{figure}

Next, we move on to probing alignments between quadrupole and higher multipoles, if any. The angular separation, $\theta_{2l}$, between $l=2$ and any higher multipole $l\geq3$
upto $l_{max}=61$ (i.e., the first sixty multipoles) are shown in the
\emph{top} panel of Fig.~[\ref{fig:algn-l2ell}]. Recall that the PEVs are headless
vectors and the maximum angular separation possible between them is $90^\circ$ (degrees).
In the \emph{bottom} panel of the Fig.~[\ref{fig:algn-l2ell}], the $p$-value of the observed level of
alignment between the quadrupole and rest of the multipoles up to $l_{max}=61$ are
shown. We find that the modes aligned with the quadrupole are $l=44$, $59$ and $61$ (beside
$l=3$) with a probability of $p\leq 0.05$, on comparison with simulations.
The $2\sigma$ CL ($p$-value $\leq 0.05$) is indicated by a horizontal dashed grey line
in the same bottom panel of Fig.~[\ref{fig:algn-l2ell}].

\subsection{Nature of collective alignments among low multipoles}

\begin{figure}[t]
\centering
\includegraphics[width=0.48\textwidth]{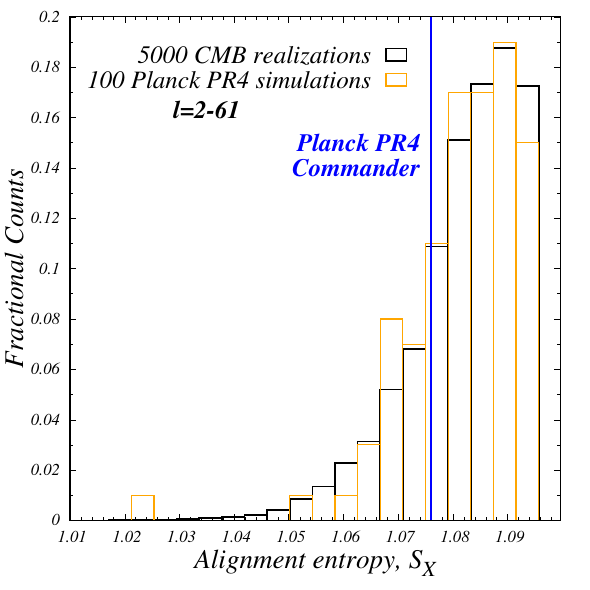}
~
\includegraphics[width=0.48\textwidth]{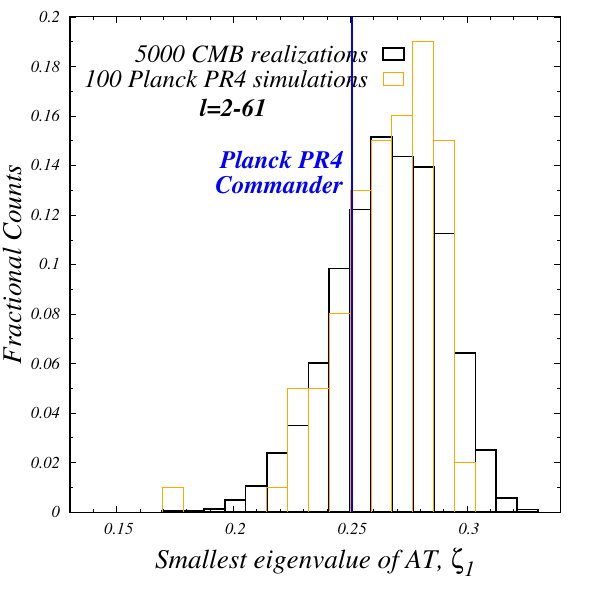}
\caption{\emph{Left :} Histogram of the Alignment entropy, $S_X$, as obtained from 100
        \planck\ PR4 \commander\ processed simulations and 5000 ideal CMB realizations generated
        based on best fit theoretical power spectrum from \planck\ 2018 data.
        The observed Alignment entropy, $S_X^{\rm obs}$, as seen in \planck\ PR4
        \commander\ CMB map for the
        range of multipole $l=[2,61]$ is indicated by the vertical blue line.
        \emph{Right :} Same as the \emph{left} figure, but for the smallest (normalized)
        eigenvalue of the Alignment tensor i.e. $\zeta_1$.}
\label{fig:ae-at-ev1}
\end{figure}

In this section, we analyze collective alignments among low multipoles, using PEVs of the CMB modes from $l$=2 to 61. We do so using the Alignment tensor (AT)
given by Eq.~(\ref{eq:at}). For this collection of axes, the Alignment entropy (Eq.~(\ref{eq:ae}))
is shown in the \emph{left} panel of Fig.~[\ref{fig:ae-at-ev1}] as a blue vertical line overlayed
on the histograms. The two histograms of Alignment entropy, $S_X$, shown are derived from the
two sets of simulations i.e., the 100 \planck\ like simulations and 5000 ideal CMB realizations
consistent with \planck\ estimated latest cosmological parameters. The observed Alignment entropy
was found to have a chance occurrence probability of $p=0.21$ and $0.2258$, respectively when using the
two - \planck\ like and ideal CMB - simulation ensembles. These two values are
fairly consistent with each other.

Now, we check for any preferred plane in which the PEVs may be lying, whose normal is given by the eigenvector corresponding to the lowest (normalized) eigenvalue of the Alignment tensor. The lowest eigenvalue of AT, denoted by $\zeta_1$, is represented by a vertical blue line in the \emph{right} panel of Fig.~[\ref{fig:ae-at-ev1}]. The corresponding histograms for $\zeta_1$ from the two sets of simulations are also plotted in the same figure, as vertical bars in different colors. We find that the $\zeta_1^{\rm obs}$ from data has a probability
of $p=0.20$ and  $0.2436$, respectively, using 100 \planck\ like and 5000 pure CMB realizations,
which again fairly agree with each other.
These two results are summarized in Table~\ref{tab:at-ae-e1}.

Thus, using the Alignment tensor method, we find that there is no collective axial or planar alignment preferred in the orientation of PEVs of the CMB multipoles from the range $l=$2 to 61 that we studied.

\begin{table}
\centering
\begin{tabular}{c c}
  & $p$-value \\
\hline
$S_X^{\rm obs}$  & 0.21 (0.2258) \\
$\zeta_1^{\rm obs}$ & 0.20 (0.2436) \\
\hline
\end{tabular}
\caption{$p$-values of the observed Alignment entropy ($S_X^{\rm obs}$) and the smallest
         eigenvalue ($\zeta_1^{obs}$),
         respectively, from the Alignment tensor are listed here. $p$-value of the data
         statistic on comparison with 100 \planck\ like simulations is quoted principally, while
         that computed from 5000 pure CMB realizations is quoted in braces.}
\label{tab:at-ae-e1}
\end{table}

\section{Discussion and Conclusions}
In this paper, we probed the status of statistical anisotropy of CMB low multipoles using the latest
\planck\ data. The CMB map used is derived from \planck\ 2020 (PR4) data that is processed using the
\npipe\ pipeline and cleaned using the \commander\ method. We employed the Power tensor method to
evaluate any deviation from isotropy of first sixty multipoles i.e., $l=[2,61]$. 
Power entropy defined using the eigenvalues of the Power tensor is used to understand any preferred axis associated with each multipole. Eigenvector corresponding to the largest eigenvalue of the Power tensor is taken to be that preferred axis and is referred to as principal eigenvector (PEV). Further, the smallest eigenvalue of the Power tensor corresponding to a multipole sheds light on planarity of that CMB mode, depending on its smallness compared to simulations.

Then, collective alignment preferences were also probed in this range of CMB modes using the Alignment tensor statistic. Alignment tensor is constructed using the principal eigenvectors (PEVs) of the Power tensor. In order to quantify our statistics, 100 \planck\ like simulations processed using the
\commander\ method that were released as part of the final public data release (PR4) from
\planck\ mission were used. Since this simulation ensemble is small in number, we also
generated 5000 mock CMB maps using the best fit theoretical temperature power spectrum ($C_l^{th}$)
that is derived based on the latest cosmological parameters from \planck.

We find that the modes $l=$6, 13, 14, 16, 17, 30, 34, 40, and 56 (a total of 9 modes) have a
Power entropy, $S(l)$, whose $p$-value is $\leq0.05$. Assuming that a threshold probability of
$\mathbb{P}=0.05$ ($2\sigma$ level) may be deemed as anomalous, the effective probability
of finding $k_* = 9$ modes or more below this threshold $p$-value out of $n=60$ multipoles we
analyzed can be estimated using the Binomial distribution as
\begin{equation}
P_{\rm cuml}(k_*=9,n=60,\mathbb{P}=0.05)  = \sum_{k=k_*}^n \Comb{n}{k} \mathbb{P}^{k}\, (1-\mathbb{P})^{n-k}\,.
\end{equation}
This cumulative probability turns out to be $P_{\rm cuml}\approx0.0028$ which
is close to a $3\sigma$ significance. So the number of low multipoles having anomalous
Power entropy i.e., modes that are intrinsically anisotropic is high.

Further when probing the planar anisotropy of CMB modes using the smallest eigenvalue of the
Power tensor, we find that the set of multipoles which are anomalous with a
$p$-value $\leq 0.05$ form a subset of the multipoles which have anomalously low Power entropy.
Hence it may be inferred that there are no truly planar modes.

A spurious alignment found between the quadrupole ($l=2$) and octopole ($l=3$) modes of the CMB sky since the release of full sky high-resolution CMB maps from NASA's WMAP probe has been studied using a variety of tests in the cosmology literature. Probing alignments among various multipoles, we find that the alignment between $l=2,3$ modes in \planck's \npipe\ \commander\ CMB map is marginally anomalous at about $2\sigma$ level
with a $p$-value of 0.08 and 0.035, respectively, using the two sets of simulation ensembles prepared for the present study. However, as mentioned, this is a conservative estimate given the tendency of foreground residuals to misalign intrinsically aligned modes.

We also probed alignment of higher multipoles with the quadrupole using PEVs.
The modes $l=$44, 59 and 61 beside $l=3$ (i.e.,
a total of 4 modes) were found to be aligned with $l=2$ at a significance of $5\%$ or less.
Employing the same Binomial statistics, we get a cumulative probability of
$P_{\rm cuml}(k_*=4,n=59,\mathbb{P}=0.05) \approx 0.34$ to find 4 or more number of modes out of the
rest of 59 multipoles ($l=[3,61]$) that are anomalously aligned with the quadrupole at a significance
of $2\sigma$ or better. Hence the significance of these (many number of) multipoles aligned with $l=2$ collectively in the range studied is low.

Finally, in order to understand any preferred orientation along an axis or in a plane for the set of multipoles under study, we employed the Alignment tensor statistic. Recall that the PEVs furnish independent information serving as an axis of anisotropy for that multipole, apart from what can be inferred from the (normalized) eigenvalues of the Power tensor. However, neither of the anisotropic patterns - a preferred axial or planar orientation - were found in the data when collective alignments were probed.

\begin{figure}
\centering
\includegraphics[width=0.75\textwidth]{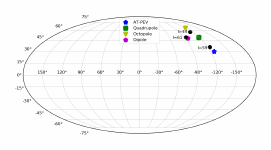}
\caption{Various anisotropy axes (PEVs) corresponding to different multipoles
	     obtained using the Power tensor in the range $l=[3,61]$ that are aligned
	     with the quadrupole ($l=2$)
         at a significance of $p\leq 5\%$ are indicated here (see section~\ref{sec:pt-algn-2l}).
         Also indicated is the Alignment tensor PEV for the entire multipole range $l=2$ to 61.
         Interestingly it is in close proximity to the anomalous multipoles whose PEV is aligned
         with the CMB quadrupole PEV at a significance of $p\leq0.05$.}
\label{fig:axes}
\end{figure}

Thus there is a hint for anomalous axial anisotropy of CMB modes at low multipoles in line with what was found in earlier studies using the Power tensor method. But no planar anisotropy of CMB modes individually or such orientation of modes collectively in a plane, of any significance, was seen in the data. Multipole PEVs that are anomalously aligned with $l=2$ with a $p$-value of 5\% or less are shown in Fig.~[\ref{fig:axes}], along with the CMB kinetic dipole \cite{Fixsen09}, and the Alignment tensor PEV (AT-PEV) for the full range of multipoles $l=2$ to 61. 
This AT-PEV is, interestingly, also close to the cluster of PEVs anomalously aligned with the quadrupole at a significance of 5\% or better.
The angular separation between them and the AT-PEV are listed in Table~\ref{tab:at-pev-l-algn}.

With upcoming ground and space based CMB experiments such as, for example,
CMB-S4~\cite{CMBS4ScienceCase2019}, Simon's Observatory~\cite{SimonsObs2019}
and Lite-BIRD~\cite{LiteBIRD2022} targeting precise measurement of CMB polarization signal on a large fraction or over full sky, it will be interesting to see how the many instances of isotropy violation seen in CMB temperature sky such as axiality, planarity and preferred alignments among its low multipoles studied here will fare as they are sourced by the same primordial fluctuations. These may be confirmed or ruled out as statistical flukes both of which will help in building a robust model of our universe. In the future, we plan to perform such a study with polarization data as is available now although its signal to noise ratio is low, and also using simulations assuming the mission specifications of these future CMB experiments.


\begin{table}
\centering
\begin{tabular}{c c}
\hline
Multipole  & Angular distance\\
`$l$'        & from AT-PEV \\
\hline
1 & $29.2^\circ$ \\
2 & $20.1^\circ$ \\
3 & $34.6^\circ$ \\
44 & $28.7^\circ$ \\
59 & $6.0^\circ$ \\
61 & $31.1^\circ$ \\
\hline
\end{tabular}
\caption{Angular separation between collective alignment axis inferred from PEVs of all sixty multipoles $l=[2,61]$ using the Alignment tensor statistic, and various low-$l$ modes anomalously aligned with CMB quadrupole PEV.}
\label{tab:at-pev-l-algn}
\end{table}

\section*{Acknowledgements}
Some of the results in the current work were derived using the publicly available
\healpix\ package \cite{healpix}.
Part of the results presented here are based on observations obtained with \planck\
(\verb+http://www.esa.int/Planck+), an ESA science mission with instruments and
contributions directly funded by ESA Member States, NASA, and Canada.
We acknowledge the use of WMAP satellite data from Legacy Archive for Microwave Background Data Analysis (LAMBDA), part of the High Energy Astrophysics Science Archive Center (HEASARC). HEASARC/LAMBDA is a service of the Astrophysics Science Division at the NASA Goddard Space Flight Center.
We also acknowledge the use of \texttt{CAMB}, a freely available Boltzmann solver for CMB anisotropies
\cite{camb1,camb2}.
The present work made use of \isap\ software \cite{isap}.
This research used resources of the National Energy Research Scientific Computing
(NERSC) Center\footnote{\url{https://www.nersc.gov/}}, which is supported by the
Office of Science of the U.S. Department of Energy under Contract No. DE-AC02-05CH11231.
Further the authors would like to thank the handling Editor of the journal M. Doser for the support extended during the long course of our manuscript's review process.


\appendix

\section{CMB maps and Masks used}\label{apdx1}
In this section, we describe the CMB temperature anisotropy maps and galactic masks used in the present study. Cleaned CMB map obtained with the \commander\ method using \planck\ mission's
complete observational data that is processed using \npipe\ pipeline was released as part of its
fourth and final public data release (PR4) \cite{Planck2020}.

\begin{figure}
\centering
\includegraphics[width=0.46\textwidth]{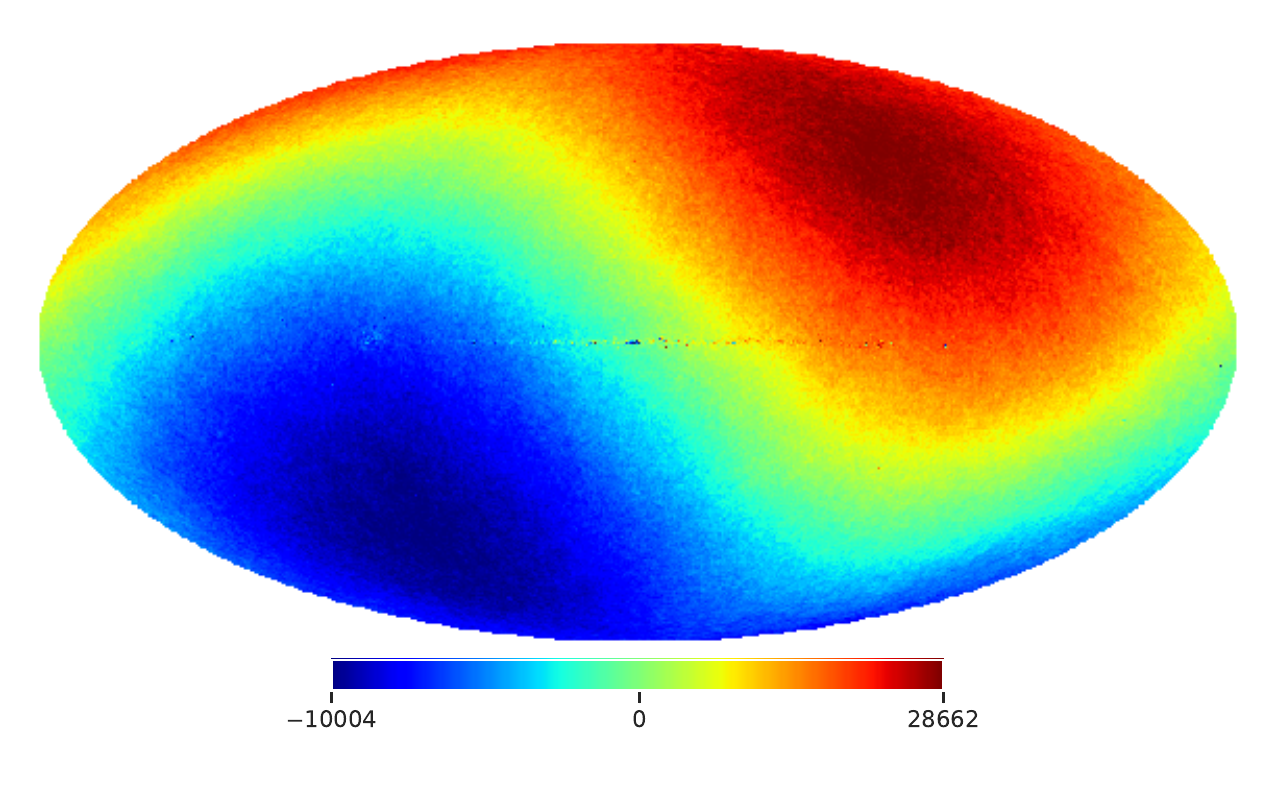}\\
\includegraphics[width=0.42\textwidth]{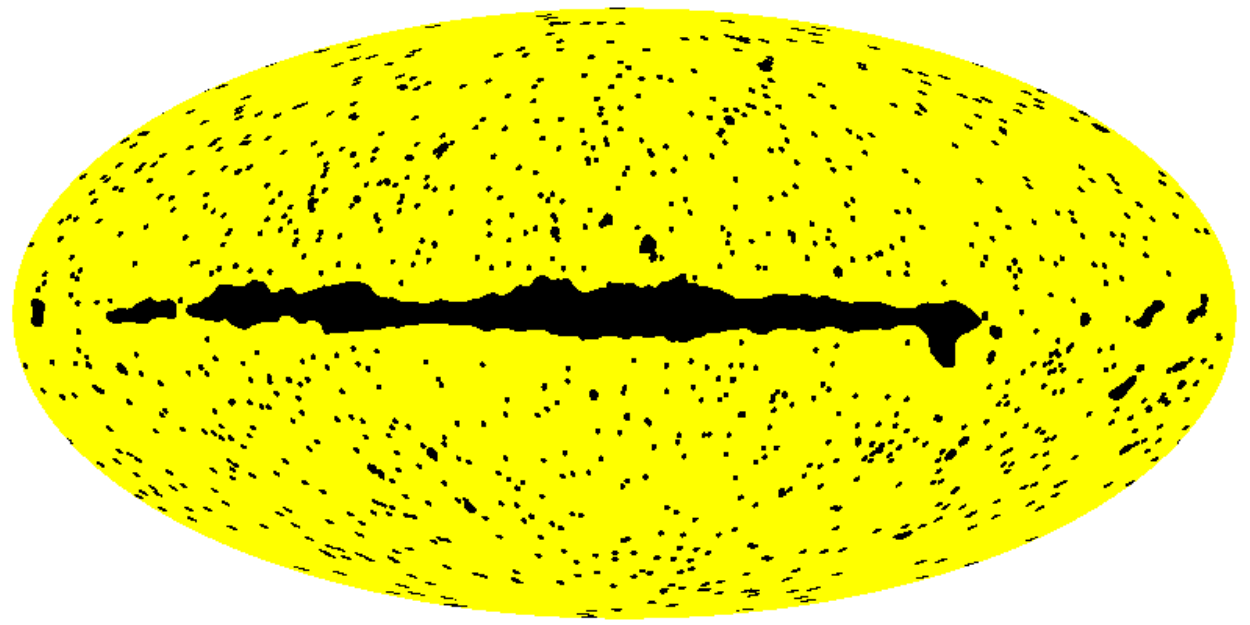}
~
\includegraphics[width=0.42\textwidth]{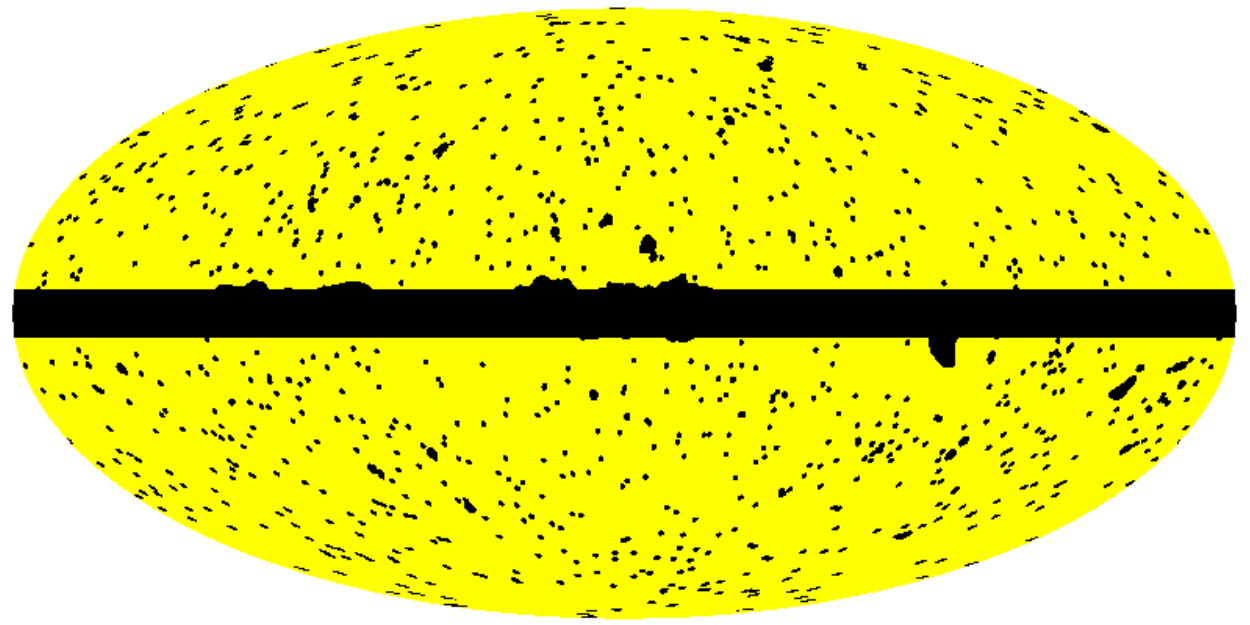}\\
\includegraphics[width=0.42\textwidth]{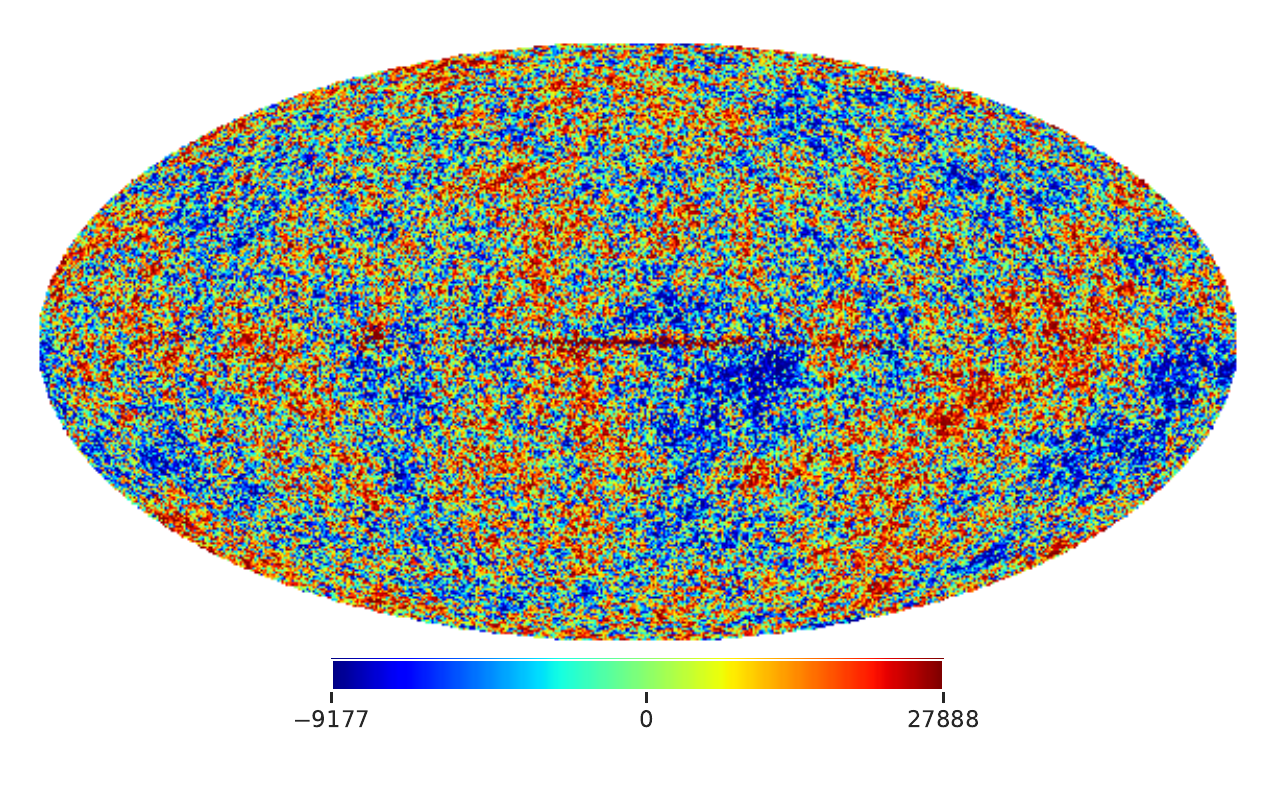}
~
\includegraphics[width=0.42\textwidth]{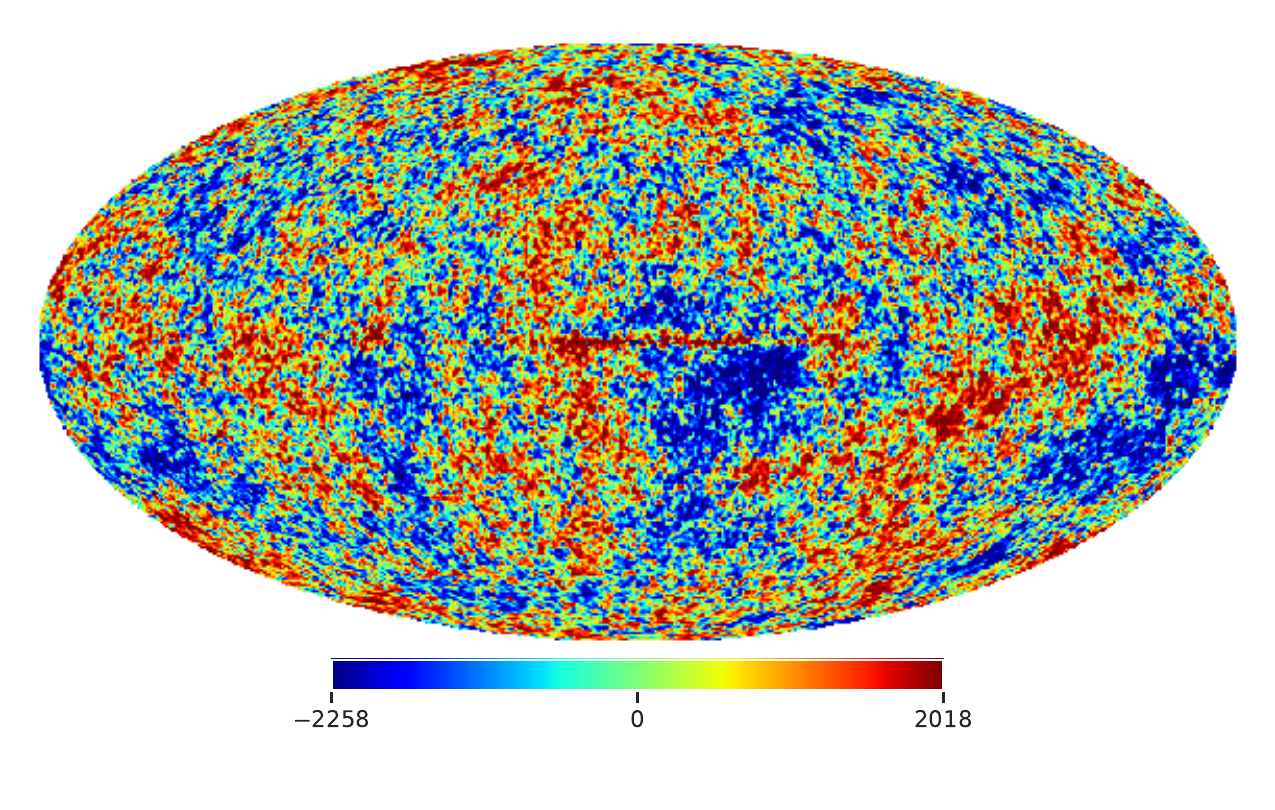}\\
\belowbaseline[0pt]{\includegraphics[width=0.42\textwidth]{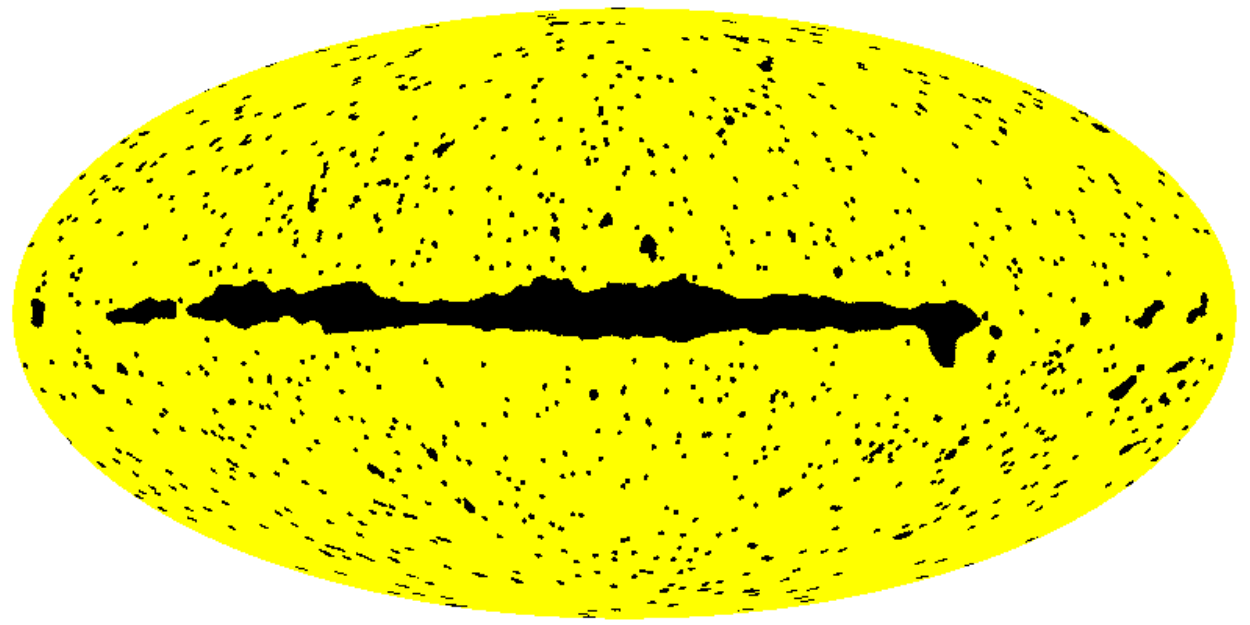}}
~
\belowbaseline[0pt]{\includegraphics[width=0.42\textwidth]{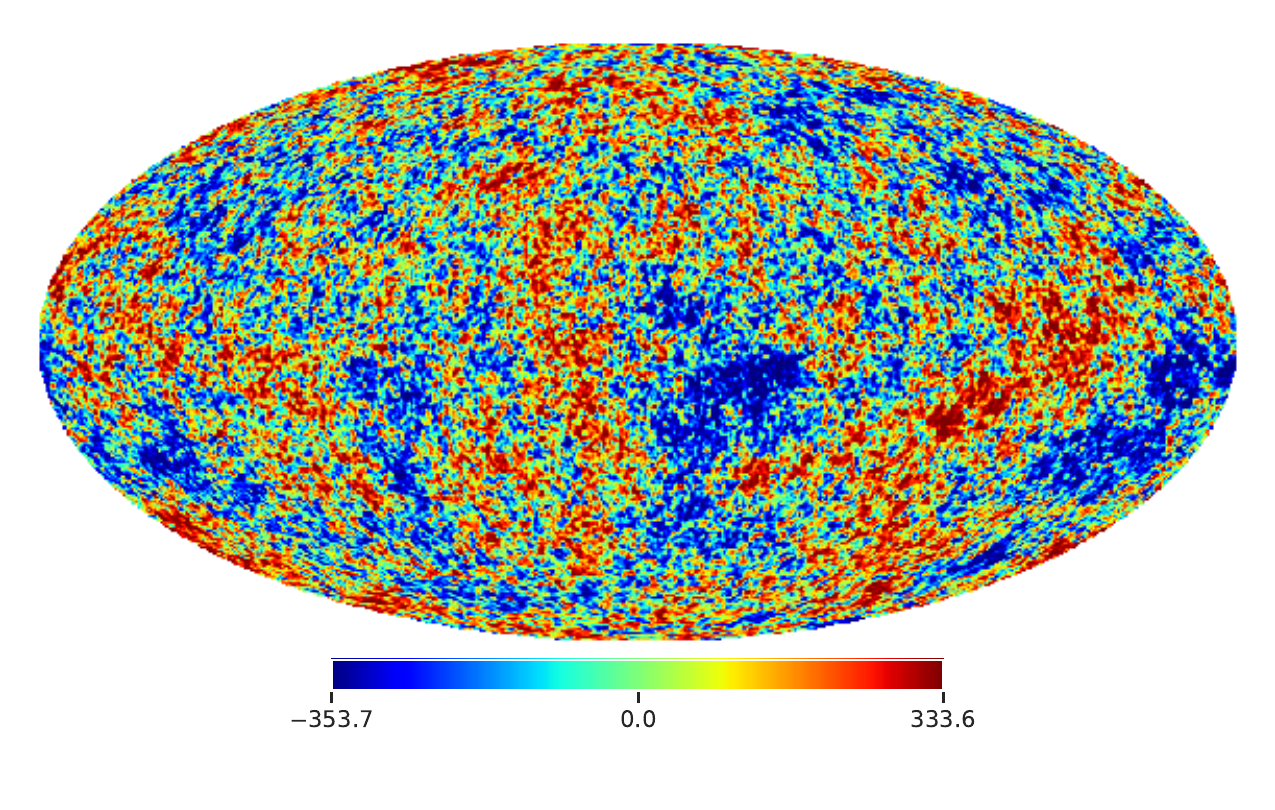}}
\caption{\emph{First row :} \commander\ cleaned CMB temperature map from \planck's public release 4.
	\emph{Second row :} Galactic mask used in the present study with a sky fraction of
	$f_{sky}\approx 0.92$
	is shown to the \emph{left} and the same mask with additional azimuthal cut to fit out the dipole
	in the CMB map from top row is shown to the \emph{right}.
	\emph{Third row :} \commander\ 2020 CMB map after fitting out the dipole is shown to the \emph{left}
	and the same map after downgrading to \healpix\ $N_{side}=256$ is shown to the \emph{right}.
	\emph{Fourth row :} The galactic mask from left of second row is originally available at
	$N_{side}=512$. It's appropriately extended to obtain a mask at $N_{side}=256$ as described
	in the text and is displayed here on the \emph{left} side. Finally, the CMB \commander\ 2020
	temperature map at $N_{side}=256$ after inpainting with \isap\ using the mask on the left side
	is displayed in the \emph{right} panel.}
\label{fig:apdx1-cmb-masks}
\end{figure}

In Fig.~[\ref{fig:apdx1-cmb-masks}], \planck's \commander\ 2020 CMB map is shown at the \emph{top} 
of the figure. Note that this map is produced at a \healpix\ resolution of $N_{side}=4096$, and
contains the dipole ($l=1$) component also.
In the \emph{second} row, galactic mask used in the present analysis is shown
in the \emph{left} panel, and the mask on the \emph{right} is the same one with additional
azimuthal cut of $\cos(b)=\pm 0.1$ along the galactic plane (where `$b$' is the galactic co-latitude)
that is used to fit-out the dipole.
The galactic mask shown in the left panel of the second row has an available sky fraction of
$f_{sky}\approx0.92$, and is obtained by extending the WMAP's nine year kp8 temperature
cleaning mask\footnote{\url{https://lambda.gsfc.nasa.gov/product/wmap/dr5/m_products.html}}
which includes point sources as well \cite{WMAP9yrFinalmaps}.

We note the following, for choosing this custom mask. Since we are interested in the low multipoles, the mask used in the inpainting procedure with \isap\ has to be chosen carefully with as much non-zero sky fraction as possible to avoid any bias in the reconstructed CMB map. We found that with an aggressive mask like the \planck\ 2018 Common mask with $f_{sky}\approx0.78$,
the inpainted map shows some visible discontinuities at the mask boundary.
\planck's \commander\ 2018 temperature mask has an $f_{sky}\approx0.88$. Finally, the common
inpainting mask used in \planck's 2018 analysis of temperature maps has an $f_{sky}\approx0.979$,
with very few point sources and a small contiguous portion in the Galactic plane. Thus we chose the extended WMAP's nine year kp8 mask with $f_{sky}\approx0.9$ that is intermediate between the aggressive Common mask on one hand and a highly conservative \emph{Planck} common inpainting mask on the other hand. This mask is originally available at \healpix\ $N_{side}=512$. First, we upgraded
it to $N_{side}=4096$ and smoothed with a Gaussian beam of $FWHM=30'$ ((arcmin) to smooth the mask boundaries up on upgradation. Then a cutoff of 0.8 was applied on the upgraded smoothed mask to get the extended mask. Few isolated regions were also excluded from the final mask that is shown in the \emph{left} panel of second row in Fig.~[\ref{fig:apdx1-cmb-masks}].

In the \emph{third} row, displayed on the \emph{left} is the \npipe\ \commander\ map (shown at the top of the figure) after fitting out the dipole using the azimuthally symmetric mask shown in the second row to the right. One can readily see the residual contamination in the galactic plane. Shown in the \emph{right} panel of \emph{third} row is the same map downgraded to $N_{side}=256$,
following Eq.~(\ref{eq:con-decon}).

In the \emph{fourth} row, shown to the \emph{left} is the downgraded WMAP's nine year kp8 mask
to $N_{side}=256$. This downgraded mask is applied to the smoothed downgraded \planck\ \commander\
2020 CMB temperature map shown in the third row to the right. The inpainted map thus obtained is displayed in the \emph{right} panel of \emph{fourth} row.

Simulations, totaling 100 maps, complementing the \planck's \commander\ 2020 CMB temperature map were also released as part of PR4. These are provided at the same resolution as data map. Hence we process viz., downgrade and inpaint the simulations also in the same way as data map. Since, the number of \planck\ provided simulations are less in number, we also generated ideal CMB realizations at
\healpix\ $N_{side}=256$ with $40'$ Gaussian beam smoothing. The histograms thus generated
from the 100 \planck\ like FFP realizations and 5000 pure CMB maps compare fairly well for the various statistics computed, as demonstrated in the paper. The larger set of ideal CMB maps provided only better characterization of our statistics i.e., in assessing their significances.


\bibliographystyle{unsrt}
\bibliography{skp1_ref_etal	}

\end{document}